\documentclass[10pt,journal,compsoc]{IEEEtran}

\ifCLASSOPTIONcompsoc
  \usepackage[nocompress]{cite}
\else
  \usepackage{cite}
\fi
\ifCLASSINFOpdf
  \usepackage[pdftex]{graphicx}
\else
\fi
\usepackage{amsmath}
\usepackage{physics} 

\usepackage[colorlinks=true,linkcolor=blue,citecolor=blue,urlcolor=blue]{hyperref}

\usepackage{ifthen}

\newboolean{VersionForJournal}
\setboolean{VersionForJournal}{false}

\newboolean{IncludeAppendix}
\setboolean{IncludeAppendix}{false}
\newboolean{UseHighResolutionImages}
\setboolean{UseHighResolutionImages}{true}

\newcommand{\HT}{\dagger}   
\newcommand{\OHT}{*}        

\newcommand{\NN}{n}  
\newcommand{\MM}{m}  
\newcommand{\DD}{d}  

\usepackage{soul} 
\usepackage{color}
\newcommand{\removedtext}[1]{} 

\usepackage{listings}
\usepackage{xcolor}
\definecolor{customblue}{RGB}{0, 176, 240}
\definecolor{custommagenta}{RGB}{255, 160, 255}
\definecolor{customgreen}{RGB}{160, 255, 160}

\usepackage[T1]{fontenc}
\usepackage{textcomp}

\lstdefinestyle{pseudocode}{
    basicstyle={\fontsize{4.4pt}{5.2pt}\selectfont\ttfamily\bfseries},
    escapechar=@,  
    mathescape=true,  
    columns=fixed,
    keepspaces=true,
}

\newcommand{\tightcolorbox}[2]{%
  {\setlength{\fboxsep}{0.3pt}%
   \setlength{\fboxrule}{0.3pt}%
   \fcolorbox{#1}{#1}{\textcolor{black}{#2}}}}

\usepackage{graphicx}   

\usepackage{stmaryrd} 

\begin{document}

\title{How to Write a Simulator for Quantum Circuits from Scratch: A Tutorial}
%
%
%
%

\author{Michael~J.~McGuffin,
        Jean-Marc Robert
        and Kazuki~Ikeda
\IEEEcompsocitemizethanks{\IEEEcompsocthanksitem M. McGuffin and J.-M. Robert are with \'{E}cole de technologie sup\'{e}rieure,
Montreal,
Canada.
\IEEEcompsocthanksitem K. Ikeda is with University of Massachusetts Boston,
Boston, USA.
}
\ifthenelse{\boolean{VersionForJournal}}{
\thanks{Manuscript received Whenuary nn, 2025; revised Whenember nn, 20nn.}
}{} 
}

%
%

\ifthenelse{\boolean{VersionForJournal}}{

\markboth{Transactions on Quantum Engineering,~Vol.~V, No.~N, Whenuary~2025}%
{McGuffin \MakeLowercase{\textit{et al.}}: How to Write a Simulator ...}
%

}{} 



\IEEEtitleabstractindextext{%
\begin{abstract}
This tutorial guides a competent programmer through the crafting of a quantum circuit simulator from scratch,
even for readers with almost no prior experience in quantum computing.
Open source simulators for quantum circuits already exist,
but a deeper understanding is gained by writing ones own.
With roughly 1000-2000 lines of code, one can simulate Hadamard, Pauli X, Y, Z, SWAP, and other quantum logic gates, with arbitrary combinations of control and anticontrol qubits, on circuits of up to 20+ qubits, with no special libraries, on a personal computer.
We explain key algorithms for a simulator, in particular:
qubit-wise multiplication for updating the state vector,
and partial trace for finding a reduced density matrix.
We also discuss optimizations,
and how to compute qubit phase, purity,
and other statistics.
A complete example implementation in JavaScript is available at https://github.com/MJMcGuffin/muqcs.js ,
which also demonstrates how to compute von Neumann entropy, concurrence (to quantify entanglement), and magic,
while remaining much smaller and easier to study than other popular software packages.
\end{abstract}

%
}

\maketitle

\IEEEdisplaynontitleabstractindextext

%
\IEEEpeerreviewmaketitle

\IEEEraisesectionheading{\section{Introduction}\label{sec:introduction}}

%
%
%
%
\IEEEPARstart{F}{or} many fields in computer science,
one road to mastery for students, practitioners, and researchers is to write their own
basic software tools,
such as a compiler (in the case of programming languages),
web server or client (for web programming),
animation or rendering engine (for computer graphics), etc.
Authoring these enables a deeper understanding of key design choices, algorithms, and tradeoffs.
For quantum computing, a circuit simulator is one such basic tool.
Several open-source implementations already exist, however,
the popular ones we checked are sizable:
Cirq, cuQuantum, Pennylane, Qiskit, and QuTiP each comprise between 40k to 500k lines of source code across hundreds of Python files,
while QuEST \cite{jones2019quest} and CUDA-Q comprise over 30k and 190k lines of C/C++ respectively.
Such large software packages are designed to offer a broad range of functionality,
but internally, they often assume expertise and involve multiple layers of abstraction,
making it difficult for newcomers to find, let alone understand, the core algorithms of interest.
They are not designed primarily as reference implementations for students
to inspect and from which to learn how to implement their own simulator.

%
%

This tutorial is designed to cover the minimum concepts necessary
to write a circuit simulator, assuming the reader has almost no
previous experience
with quantum computing, but does have experience programming,
and has an understanding of basic linear algebra and algorithm complexity
(big-O notation).
We introduce basics such as bra-ket notation and tensor (Kronecker) products,
but rather than using abstract, mathematical definitions,
we prefer concrete, informal, computational explanations
that are likely to be easier to understand by engineers and software developers.
We then discuss in more detail two key algorithms:
{\em qubit-wise multiplication}, which performs efficient updating of a state vector,
and {\em partial trace}, which is necessary to compute reduced density matrices
that can then be used to
analyze a subsystem of a larger quantum system.
Unlike previous resources (e.g., \cite{viamontes2009qcsim,maziero2017computing,xu2023}),
we provide graphical explanations of these algorithms,
with well-commented pseudocode written in the style of a modern programming language,
and also provide an open source example implementation \cite{mcguffin2025muqcs} in JavaScript,
a language chosen so it can be executed in a web browser's console without installing any extra software.
The implementation also contains subroutines for computing
qubit phase, purity, von Neumann entropy,
concurrence \cite{coffman2000distributed} (to quantify the entanglement of a pair of qubits),
and stabilizer R\'{e}nyi entropy \cite{leone2022renyi} (to quantify magic),
all in well under 3k lines of code.
A companion video\footnote{\url{https://youtu.be/b6OqXkqPBeY}} gives an overview of the source code in the example implementation,
pointing out key parts of it,
and demonstrates how to run it.

\section{Basic Concepts}\label{sec:basics}


{\em Qubits} are quantum bits, and can store a 0 and 1 simultaneously in a superposition, which is collapsed when a measurement is performed, resulting in a 0 or 1 with some probability.
The state $| \psi \rangle$ of a single qubit can be described using two complex numbers, $a_0$ and $a_1$,
where their squared magnitudes $|a_0|^2$ and $|a_1|^2$ are the probabilities\footnote{This is formalized in the {\em Born rule}.}
of measuring a 0 and 1, respectively, and
$|a_0|^2 + |a_1|^2 = 1$.
These complex numbers are called {\em amplitudes}, and are often packaged together as elements in a $2 \times 1$ column vector written as
$\ket{\psi} = \scalebox{0.8}{$\begin{bmatrix} a_0 \\ a_1 \end{bmatrix}$}$.
In the case of a single qubit,
there are two (computational) {\em basis} states,
defined as
$\ket{0} = \scalebox{0.8}{$\begin{bmatrix} 1 \\ 0 \end{bmatrix}$}$ and
$\ket{1} = \scalebox{0.8}{$\begin{bmatrix} 0 \\ 1 \end{bmatrix}$}$.
Thus $\ket{\psi} = a_0 \ket{0} + a_1 \ket{1}$,
and $\ket{\psi}$
is a normalized vector in a two-dimensional complex vector space that can be expressed as a unique linear combination
of the two vectors forming a basis of that vector space.

The Dirac notation (or bra-ket notation, from the word ``bracket'') writes a row vector $u$ as $\langle u |$ (read as ``bra u'') and a column vector $v$ as $| v \rangle$ (``ket v''),
defined so that $\langle u | = | u \rangle^\HT$, where the dagger $^\HT$ denotes the {\em conjugate transpose}\footnote{This is also called the Hermitian conjugate or Hermitian transpose, sometimes denoted with a $^\OHT$ symbol.}, i.e., the transpose of a matrix or vector where every complex element $a$ is replaced with its conjugate $a^{*}$ 
(negating the imaginary component).
For example,
if $\ket{v} = \scalebox{0.8}{$\begin{bmatrix} q+ir \\ s+it \end{bmatrix}$}$,
where $q$, $r$, $s$, $t$ are real and $i = \sqrt{-1}$,
then $\bra{v} = \ket{v}^\HT = \begin{bmatrix} q-ir & s-it \end{bmatrix}$.

In general, if two vectors $\langle u|$ and $|v \rangle$ are both of length $d$,
then their dot product $\langle u | \cdot | v \rangle$ is abbreviated as
$\braket{u}{v}$ and yields a single complex number,
and their outer product $\ketbra{v}{u}$
yields a $d \times d$ complex matrix.


Next, we consider how to describe
a set of $\NN=3$ qubits,
which may be in a superposition.
In general, it is not sufficient to model each qubit with two amplitudes,
because there may be dependencies between qubits (due to {\em entanglement}, discussed later).
Instead, we describe the state of the qubits using
a $2^\NN \times 1$ = $8 \times 1$ column vector $| \psi \rangle$
called the state vector\footnote{Quantum states that can be described completely by such a state vector are called {\em pure} states.
There also exist {\em mixed} states, which cannot be described by a state vector, discussed in section~\ref{sec:partialTrace}.
Both pure and mixed states can, however, be described by a $2^\NN \times 2^\NN$ density matrix.
}.
Each of the 8 amplitudes $a_0$, $a_1$, ..., $a_7$ in $| \psi \rangle$ corresponds to a basis state $| 000 \rangle$, $| 001 \rangle$, ..., $| 111 \rangle$, respectively,
so $| \psi \rangle = a_0 | 000 \rangle + \ldots + a_7 | 111 \rangle$.
The basis state $|011 \rangle$, as an example, is an $8 \times 1$ column vector
with a 1 as its 4th element and zeros everywhere else.
Each amplitude $a_j$ determines the probability $| a_j |^2$ of measuring the $j$th basis state,
and each $a_j$ also has an associated phase (sometimes called the {\em argument} of the complex number).
If $\theta_j$ is the phase of $a_j$, we can express the amplitude as $a_j = |a_j|e^{i \theta_j}$.

The set of vectors
$\{ | 000 \rangle$, $| 001 \rangle$, ..., $| 111 \rangle \}$
is called the {\em computational} basis,
and spans a complex vector space
called {\em Hilbert space}, with $2^\NN = 8$ complex dimensions.
It is sometimes useful to consider alternative bases of the same Hilbert space,
but we only need the computational basis for our purposes.

The reader can check that
$\braket{\psi}{\psi} = \sum_j a_j^{*} a_j$ yields the sum of probabilities $\sum_j |a_j|^2$, which must be equal to 1.
The reader can also check that $| \psi \rangle \langle \psi |$ is
an $8 \times 8$ matrix, called the {\em density matrix},
where the diagonal elements are the real-valued probabilities.
The density matrix has the property of being {\em Hermitian},
i.e., it is equal to its own conjugate transpose.
Also, let $\Tr(M)$ denote the trace (sum of diagonal elements) of a matrix $M$,
and notice that $\Tr(| \psi \rangle \langle \psi |) = \langle \psi | \psi \rangle$
(one way to see this is to notice that
$\Tr(\ketbra{\psi}{\psi}) = \Tr(\braket{\psi}{\psi})$ by the `cycling property' of the trace,
and $\Tr(\braket{\psi}{\psi}) = \braket{\psi}{\psi}$ since the matrix is $1\times1$).

We might ask how many degrees of freedom are embodied in $\NN=3$ qubits.
Since each amplitude has two components (one real, one imaginary),
or alternatively, since each amplitude encodes a probability and a phase,
one might suppose that $| \psi \rangle$
encompasses $2^{\NN+1}$ real-valued degrees of freedom.
However, because the probabilities must sum to 1,
we have the constraint $|a_0|^2 + \ldots + |a_7|^2 = 1$.
Furthermore, rotating all amplitudes by the same angle in the complex plane
changes what is called the {\em global phase}, and has no physically measurable effect.
This arguably leaves only $(2^{\NN+1}-2)$ real-valued degrees of freedom.
In the case of a single qubit ($\NN=1$), this yields 2 degrees of freedom,
which are mapped to the surface of the Bloch sphere, introduced later in this section.

The {\em Kronecker product} (often informally called the {\em tensor product}\footnote{The Kronecker product
is defined on matrices,
while the tensor product is defined on linear maps,
which can be represented by matrices once a basis of the underlying vector space has been selected.})
constructs a larger matrix
from two smaller matrices.
The Kronecker product of a $q\times r$ matrix with an $s\times t$ matrix
produces a $qs \times rt$ matrix.
An example appears below.
One can imagine the first matrix being stretched to fill the result,
while the second matrix is copied repeatedly,
and the elements in the resulting matrix are products of elements of the first two matrices.


\noindent\includegraphics[width=0.99\columnwidth]{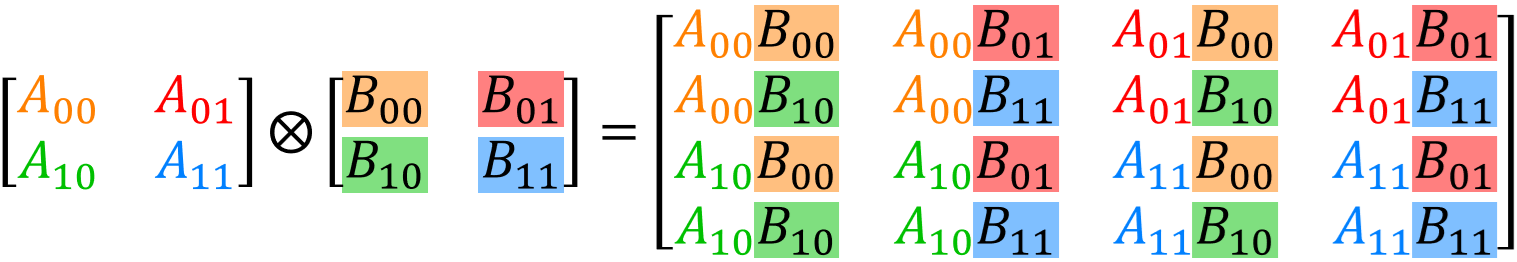}


As an exercise, the reader might check that
$| 0 \rangle \otimes | 0 \rangle \otimes | 0 \rangle$
yields an $8 \times 1$ column vector equal to $| 000 \rangle$,
and similarly,
$| 0 \rangle \otimes | 1 \rangle \otimes | 1 \rangle = | 011 \rangle$.
Applying the tensor product multiple times to the same matrix or vector
can be denoted with a variant of exponential notation, e.g.,
$| 1 \rangle ^ {\otimes 3} = | 111 \rangle$.

\begin{figure}[!thb]
\centering
    \includegraphics[width=0.99\columnwidth]{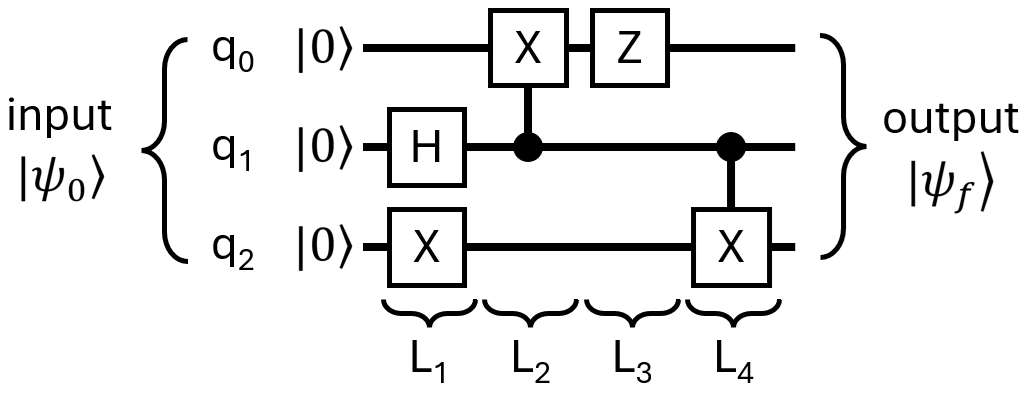}
\caption{An example circuit. Time advances left-to-right.
The 3 qubits are initially zero. The Hadamard $H$ gate shifts a qubit from a zero state $| 0 \rangle$ to an equal superposition of $| 0 \rangle$ and $| 1 \rangle$. 
The Pauli $X$ gate performs a bit flip, and the Pauli $Z$ gate performs a phase flip.
The large dots vertically connected to two $X$ gates indicate {\em control qubits},
and they form ``controlled $X$'' or $\mathrm{CX}_{j,k}$ gates,
which flip the {\em target qubit} $q_k$ if the control qubit $q_j$ is $\ket{1}$.
The circuit transforms the initial state $\ket{\psi_0} = \ket{0} \otimes \ket{0} \otimes \ket{0}$
to the final state $\ket{\psi_f}$,
and the text shows how to calculate that $\ket{\psi_f} = (1/\sqrt{2})(\ket{100}-\ket{011})$.
}
\label{fig:exampleCircuit}
\end{figure}

Figure~\ref{fig:exampleCircuit} shows a quantum circuit
on 3 qubits $q_0$, $q_1$, $q_2$.
We adopt the convention that qubits are numbered $q_0$ to $q_{n-1}$
increasing from top to bottom in circuit diagrams,
and decreasing from left to right\footnote{\label{fn:ordering}This
matches the convention in multiple software packages \cite{gidney2017blog,gidney2017quirkReadme,freeze2022,ibm2024ordering,ibm2025ordering}
but is opposite the convention in many physics textbooks.}
in tensor products,
e.g., $\ket{\psi} = q_2 \otimes q_1 \otimes q_0$,
which makes sense if we imagine $q_{n-1}$ storing the most significant bit,
or high-order bit, of a multi-bit binary number.


The boxes in Figure~\ref{fig:exampleCircuit} containing letters $H$, $X$ and $Z$
are quantum logic gates:
$H$ for a Hadamard gate,
$X$ for Pauli X gate (also called a NOT gate) and $Z$ for Pauli Z.
In the figure, two of the $X$ gates are connected vertically
to control qubits;
these are called controlled $X$ gates, or CX, or CNOT gates.
CX gates only have an effect
when the qubit at the control qubit is on.
If a control qubit is in a superposition of on and off,
then the gate outputs a superposition of having an effect
and not having an effect on the target qubit.

To ease reading, we sometimes add subscripts to the gate matrices to specify which qubit is being acted on.
For example, in Figure~\ref{fig:exampleCircuit}, there is an $H_1$ acting on qubit $q_1$ and a $Z_0$ acting on qubit $q_0$.
Initially, in that circuit, all three qubits are off (in a $\ket{0}$ state).
In the first layer $L_1$,
the $H_1$ gate shifts $q_1$ into a superposition of $\ket{0}$ and $\ket{1}$,
and the $X_2$ gate in the same layer inverts $q_2$ to a state of $\ket{1}$.
In the next layer $L_2$,
there is a $\mathrm{CX}_{1,0}$ gate (the subscripts indicate control and target qubits, respectively),
and since the control qubit is in superposition,
$q_0$ will be both flipped and not flipped,
causing qubits $(q_1,q_0)$
to enter a superposition of $|00 \rangle$ and $|11 \rangle$
(this is an example of a Bell state, where the qubits are entangled).
In other words, after layer $L_2$, we have a superposition of
$(q_2,q_1,q_0) = \ket{100}$ and $(q_2,q_1,q_0) = \ket{111}$
with equal probability and phase, so the corresponding amplitudes are $a_{100} = 1/\sqrt{2}$ and $a_{111} = 1/\sqrt{2}$,
with all other amplitudes equal to zero.
Next, the $Z_0$ gate flips the phase of all nonzero amplitudes where $q_0 = \ket{1}$,
which changes $a_{111}$ to $-1/\sqrt{2}$, but leaves $a_{100}$ unchanged.
Finally, the $\mathrm{CX}_{1,2}$ gate, whose control qubit $q_1$ is still in a superposition,
causes its target qubit $q_2$ to be both flipped and not flipped,
and the resulting nonzero amplitudes are
$a_{100} = 1/\sqrt{2}$ (unchanged since $L_2$) and $a_{011} = -1/\sqrt{2}$.
In other words, the final state is $\ket{\psi_f} = (1/\sqrt{2})(\ket{100}-\ket{011})$.
Measuring the output of the circuit causes a random collapse to
011 or 100, each with 50\% probability.

Physically, the qubits might be implemented using photons,
superconducting transmon qubits, ions, quantum dots,
etc.\ (dozens of approaches are listed in \cite{dyakonov2020}).
These details are abstracted away in a circuit diagram.
The horizontal lines in a circuit diagram are often called {\em wires},
but they do not correspond to physical wires:
instead, they represent the qubits extended over time.
The quantum gates, shown as boxes, are also not physical objects:
they represent operations performed on the qubits, somewhat like music notes in sheet music.
Also note that, in a physical quantum computer, we cannot directly measure
a superposition;
each measurement can only detect a collapsed state of 0s and/or 1s.
Thus, a circuit on a quantum computer is typically run hundreds of times, to allow indirect measurement of the probabilities.
A circuit simulator running on classical hardware, on the other hand, is exponentially slower than quantum hardware, but can explicitly compute the complete state of all qubits at all layers in the circuit, with just one simulated execution.

Each gate in a quantum circuit has a corresponding matrix.
Single-qubit gates each correspond to a $2 \times 2$ matrix.
Examples include the identity, Hadamard, and Pauli ($X$, $Y$, $Z$) gates:
\begin{align*}
I = \left[ \begin{array}{@{}cc@{}} 1 & 0  \\ 0 & 1 \end{array} \right] && H = \frac{1}{\sqrt{2}} \left[ \begin{array}{@{}cc@{}} 1 & 1  \\ 1 & -1 \end{array} \right]
\end{align*}

\begin{align*}
X = \left[ \begin{array}{@{}cc@{}} 0 & 1  \\ 1 & 0 \end{array} \right] && Y = \left[ \begin{array}{@{}cc@{}} 0 & -i \\ i & 0 \end{array} \right] && Z = \left[ \begin{array}{@{}cc@{}} 1 & 0  \\ 0 & -1 \end{array} \right]
\end{align*}

As with almost all quantum gates\footnote{Measurement gates are a notable exception.},
each of these matrices is {\em unitary}, meaning its inverse is equal to its conjugate transpose
($M^{-1} = M^\HT$).
The example matrices above also happen to be Hermitian, although this is not generally true
of gate matrices\footnote{More complete lists of gate matrices are available \cite{wikipedia2025gates,crooks2024gates,mcguffin2025muqcs}}.
A matrix that is both unitary and Hermitian is equal to its own inverse,
hence applying the gate twice cancels its effect
(e.g., the reader can confirm that $X^2 = I$ and $H^2 = I$).

To simulate a circuit such as the one in Figure~\ref{fig:exampleCircuit},
we can compute a matrix $L_j$
associated with
each of the 4 layers of the circuit.
Each $L_j$ is formed from a tensor product of the gates on each qubit (or wire)
in layer $j$.
For example\footnote{Tensor products are ordered by decreasing bit significance.
See footnote \ref{fn:ordering}.},
in Figure~\ref{fig:exampleCircuit}, $L_1 = X_2 \otimes H_1 \otimes I_0$
and $L_3 = I_2 \otimes I_1 \otimes Z_0$.
To find $L_2$ and $L_4$, we note that the CX gates correspond to 4$\times$4 matrices:
\begin{align*}
\mathrm{CX}_{j,j-1} = \scalebox{0.8}{$\left[ \begin{array}{@{}cccc@{}} 1 & 0 & 0 & 0  \\ 0 & 1 & 0 & 0 \\ 0 & 0 & 0 & 1  \\ 0 & 0 & 1 & 0 \end{array} \right]$}, && \mathrm{CX}_{j,j+1} = \scalebox{0.8}{$\left[ \begin{array}{@{}cccc@{}} 1 & 0 & 0 & 0  \\ 0 & 0 & 0 & 1 \\ 0 & 0 & 1 & 0  \\ 0 & 1 & 0 & 0 \end{array} \right]$}
\end{align*}
%
%
Hence, $L_2 = I_2 \otimes \mathrm{CX}_{1,0}$, and $L_4 = \mathrm{CX}_{1,2} \otimes I_0$.
Notice that each of the four $L_j$ matrices is unitary and $8 \times 8$.
The input to the circuit is $\ket{\psi_0} = \ket{0} ^ {\otimes 3} = \ket{000}$,
and the final output can be computed as the product $\ket{\psi_f} = L_4 L_3 L_2 L_1 \ket{\psi_0}$,
yielding an $8 \times 1$ column vector.
The reader can check that this output state is
$\ket{\psi_f} = (1/\sqrt{2})(\ket{100} - \ket{011})$.

This method for computing a circuit's output can be applied generally
to other circuits if the matrices of the gates are known.
However, the method does not scale well.
Consider a circuit over $\NN$ qubits with depth $\DD$, i.e.,
the circuit has layers $1, \ldots, \DD$.
Each layer's matrix $L_j$ will be $2^\NN \times 2^\NN$,
requiring $O(4^\NN)$ memory. Multiplying two such matrices together is also quite expensive
($O(8^\NN)$ time, using a naive algorithm).
Fortunately, we can avoid ever multiplying
two $2^\NN \times 2^\NN$ matrices
by noticing that
the right hand side of
$| \psi_f \rangle = L_\DD \ldots L_1 \ket{\psi_0}$
can be evaluated right-to-left.
Working from the right,
each product of $L_j$ with a column vector yields another column vector,
and each such product costs $O(4^\NN)$ time,
for a total time of $O(4^\NN \DD)$.


It is straightforward to prove that, given two unitary matrices such as $U_1$ and $U_2$,
their product $U_2 U_1$ is also unitary,
and therefore if all the layers $L_j$ of a circuit are unitary,
then the entire circuit $L_\DD \ldots L_1$ is unitary.
The fact that the gate matrices, and $L_j$ matrices, and the entire circuit are unitary
is related to a postulate of quantum mechanics (section 2.2.2 in \cite{nielsen2010}).
We briefly illustrate why this unitarity is important.
In Hilbert space, the squared length or squared norm of a vector $\ket{\psi}$
is defined as $\ket{\psi}^\HT \ket{\psi} = \braket{\psi}{\psi}$.
If the vector is transformed by a unitary matrix $U$,
the resulting vector is $U \ket{\psi}$,
and the squared norm of this new vector is
$(U \ket{\psi})^\HT (U \ket{\psi}) = (\ket{\psi}^\HT U^\HT)(U \ket{\psi}) = \bra{\psi} U^{-1} U \ket{\psi} = \braket{\psi}{\psi}$.
In other words, the transformation $U$ preserves the square of the norm of the vector.
But the square of the norm of a state vector is also equal to the sum of the squared
magnitudes of the amplitudes, i.e., it is equal to the sum of the probabilities.
So unitary matrices preserve the sum of probabilities associated with a state vector,
ensuring that this sum remains equal to 1.
This is sometimes called {\em conservation of probability}.
Another way to understand unitary matrices is that they rotate a state vector in Hilbert space around the origin.
Notice also that, because unitary matrices always have an inverse matrix,
it is always possible to undo the effect of a unitary matrix,
e.g., reversing the effect of a circuit.
This is an example of {\em conservation of information}.

%
%

\begin{figure}[!thb]
\centering
\includegraphics[width=0.99\columnwidth]{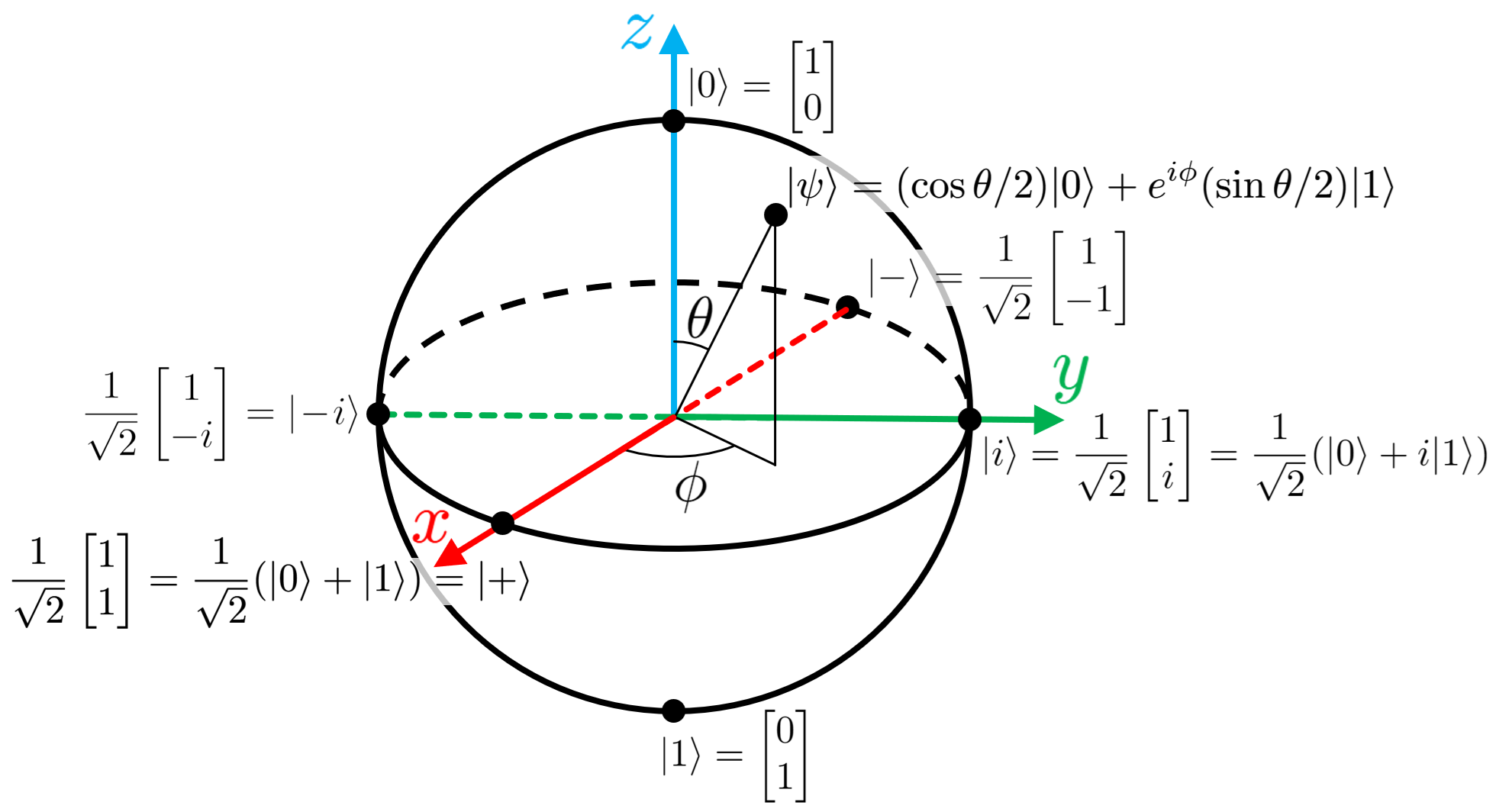}
\caption{The Bloch sphere's surface contains all physically distinguishable states of a single qubit
whose state can be described with a state vector $| \psi \rangle$.
Certain states have standard names like $|+ \rangle$ and $|i \rangle$.
}
\label{fig:bloch}
\end{figure}

To better understand the effect of any given gate, we again consider a single qubit in isolation,
described by
$\ket{\psi} = a_0 \ket{0} + a_1 \ket{1}$.
Let $\theta_0$, $\theta_1$ be the phases of $a_0$, $a_1$, respectively,
and let $\phi = \theta_1 - \theta_0$.
Then we can rewrite the state vector as
$\ket{\psi} = |a_0|e^{i \theta_0} \ket{0} + |a_1|e^{i \theta_1} \ket{1} = e^{i \theta_0} (|a_0|\ket{0} + |a_1|e^{i \phi}\ket{1})$.
The factor of $e^{i \theta_0}$ corresponds to a global phase,
and changing the angle in this factor rotates both amplitudes by the same amount,
resulting in no physically detectable change.
Thus, the state vector is physically identical to
$\ket{\psi} = |a_0|\ket{0} + |a_1|e^{i \phi}\ket{1}$.
Let $\theta = 2 \arccos|a_0|$,
hence $|a_0| = \cos(\theta/2)$.
Recall that $|a_0|^2+|a_1|^2=1$,
so similarly $\cos^2(\theta/2)+\sin^2(\theta/2)=1$,
and $|a_1| = \sin(\theta/2)$.
The state vector can then be written
$\ket{\psi} = \cos(\theta/2)\ket{0} + e^{i \phi}\sin(\theta/2)\ket{1}$.
The two angles $\theta$ and $\phi$
are one way to describe the two degrees of freedom of the state $\ket{\psi}$,
and they determine a point on the surface of the {\em Bloch sphere} (Figure~\ref{fig:bloch}).

It turns out that the effect of each single-qubit gate
is to {\em rotate} around an axis of the Bloch sphere.
For example, the $X$ gate rotates the $|0\rangle$ state to the $|1\rangle$ state,
i.e., $X |0\rangle = |1\rangle$, and more generally, the $X$ gate
rotates any state $180^{\circ}$
around the $x$-axis
in Figure~\ref{fig:bloch}.
Similarly, the $Y$ and $Z$ gates
rotate $180^{\circ}$
around the $y$- and $z$-axes,
respectively, of the Bloch sphere.
The Hadamard $H$ gate
rotates $180^{\circ}$
around the $(x+z)$-axis,
mapping $\ket{0}$ to $\ket{+} = (1/\sqrt{2})(|0 \rangle + |1 \rangle)$.
As another example, the $T$ gate, which is equal to $Z^{0.25}$
and whose matrix the reader can look up,
rotates $45^{\circ}$
around the $z$-axis.

\section{Qubit-Wise Multiplication}

\begin{figure*}[!t]
 \centering
  \ifthenelse{\boolean{UseHighResolutionImages}}{
    \includegraphics[width=1.99\columnwidth]{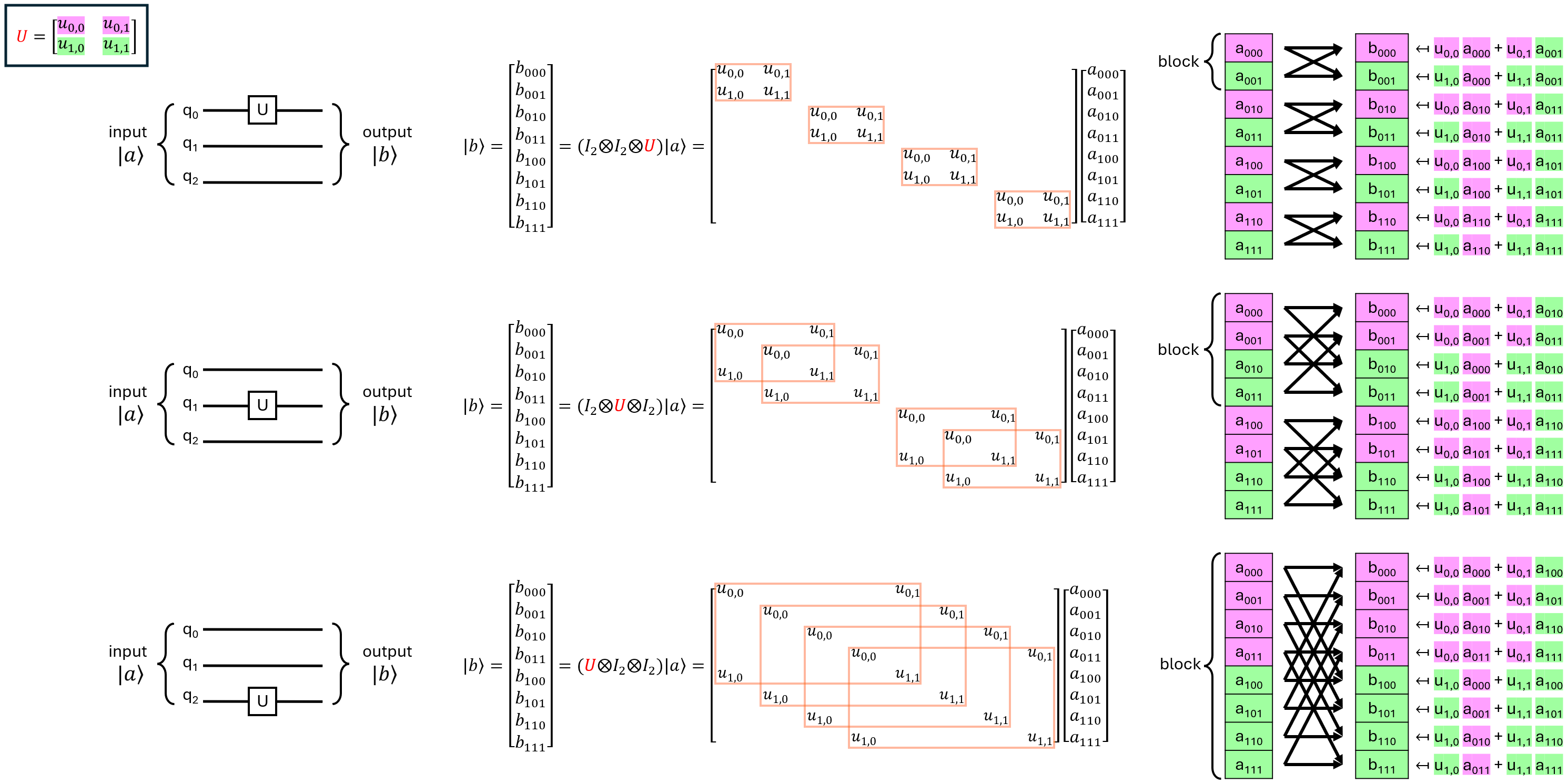}
  }{
    \includegraphics[width=1.99\columnwidth]{fig/qubitWise-lowResolution.png}
  } 
  \caption{
     The effect of a single-qubit gate, with 2$\times$2 matrix $U$,
     on an 8$\times$1 input state vector $| a \rangle$,
     can be understood by studying the patterns in how amplitudes
     are recombined to produce an output state vector $| b \rangle$.
     Upper left: definition of the $U$ matrix.
     Left column: different positions of the $U$ gate.
     Middle: $| b \rangle$ is equal to a sparse 8$\times$8 matrix
        multiplied by $| a \rangle$.
        Elements equal to zero in each 8$\times$8 matrix are omitted for clarity.
	Orange rectangles hint at repeating patterns that are instructive to compare
	to those in partial trace (Figure~\ref{fig:partialTrace}).
     Right: arrows show which amplitudes in $| a \rangle$ contribute
        to those in $| b \rangle$,
        and assignments ($\mapsfrom$) show how the elements of $U$ act like weights.
     State vectors are partitioned into even (purple) and odd (green) subsets.
     The qubit-wise multiplication algorithm computes
     $| b \rangle$ from $| a \rangle$
     without explicitly
     storing any 8$\times$8 matrix.
  }
 \label{fig:qubitWise}
\end{figure*}

Here we explain how to efficiently simulate single-qubit gates (modeled with 2$\times$2 matrices),
with optional, arbitrary combinations of control and anti-control qubits.

Readers who compute a few $L_j$ for various circuits might notice that they are often sparse, especially in layers where there is only one gate.
This is a hint that we can achieve much better time and memory performance by not explicitly storing the full matrices.
This is precisely what is done by {\em qubit-wise multiplication},
an algorithm described by Viamontes et al.\ \cite{viamontes2009qcsim}\footnote{Unfortunately, the pseudocode in their book contains some minor mistakes, and also does not allow for control qubits.}
and which we adapted from the source code for Quirk \cite{gidney2020quirk}.

The algorithm can be understood by writing out a few examples of $L_j$ matrices for layers of a circuit where there is a single 1-qubit gate.
Figure~\ref{fig:qubitWise} and the pseudocode below both refer to a generic gate matrix $U$ (which, in practice, could be $H$, $X$ or some other matrix),
and they both partition the input and output state vectors into contiguous ``blocks'', each of which has an even (purple) and odd (green) half-block.
Figure~\ref{fig:qubitWise} also illustrates the patterns in the non-zero elements that appear in the $L_j$ matrices,
and in the relationships between input and output blocks.
The pseudocode steps through the blocks (line 43),
and within each block, it steps through matching amplitudes of the even and odd half-blocks.


%
% ...


\begin{lstlisting}[style=pseudocode, basicstyle={\fontsize{5.3pt}{6.4pt}\selectfont\ttfamily\bfseries}]
01 // Returns the product of (I@$\otimes$@... $\otimes$I$\otimes$U$\otimes$I$\otimes$...$\otimes$I) and |a>,
02 // where I is the 2@$\times$@2 identity matrix, U is a given 2@$\times$@2 matrix,
03 // |a> is a (2^n)@$\times$@1 column vector, and the return value is
04 // another column vector of the same size as |a>.
05 // The tensor product in parentheses has n factors, and would
06 // result in a matrix of size (2^n)@$\times$@(2^n) if evaluated explicitly.
07 // U is at a position in the tensor product given by i_w,
08 // with i_w=0 or i_w=n-1 indicating that U 
09 // is the right-most or left-most factor, respectively.
10 // The algorithm avoids explicitly computing the tensor product
11 // in parentheses, and takes O(2^n) time.
12 @\textcolor{customblue}{// Control bits and anti-control bits limit the effect of U }@
13 @\textcolor{customblue}{// to a subset of the amplitudes in |a>. }@
14 qubitWiseMultiply(
15     n,   // number of qubits in the circuit, 1 <= n
16     U,   // a 2@$\times$@2 matrix of complex numbers
17     i_w, // index of wire on which to apply U, 0 <= i_w <= n-1
18
19     // This is the state vector to transform;
20     // a (2^n)@$\times$@1 column vector of complex amplitudes
21     a,
22 
23     @\textcolor{customblue}{// A list of pairs of the form [wire\_index, flag] where 0<=wire\_index<n}@
24     @\textcolor{customblue}{// and flag is true for a control bit and false for an anti-control bit}@
25     @\textcolor{customblue}{listOfControlBits = [ ] // empty by default}@
26 ) {
27     @\textcolor{customblue}{inclusionMask = 0;}@
28     @\textcolor{customblue}{desiredValueMask = 0;}@
29     @\textcolor{customblue}{for ( iterator : listOfControlBits ) \char123 }@
30         @\textcolor{customblue}{[ wireIndex, flag ] = iterator;}@
31         @\textcolor{customblue}{bit = 1 \hspace{0pt}\char60\hspace{0pt}\char60\hspace{0pt} wireIndex; // 2\char94 wireIndex }@
32         @\textcolor{customblue}{inclusionMask |= bit; // turn on the bit}@
33         @\textcolor{customblue}{if ( flag )}@
34             @\textcolor{customblue}{desiredValueMask |= bit; // turn on the bit}@
35     @\textcolor{customblue}{ \char125 }@
36 
37     sizeOfStateVector = 1 << n; // 2^n; could be 2, 4, 8...
38     sizeOfHalfBlock = 1 << i_w; // could be 1, 2, 4...
39     sizeOfBlock = sizeOfHalfBlock << 1; // could be 2, 4, 8...
40     b = a.copy(); // copies all amplitudes from a to b
41     // b0 is the index of the start of the block;
42     // offset is an offset within the block
43     for ( b0 = 0; b0 < sizeOfStateVector; b0 += sizeOfBlock ) {
44         for ( offset = 0; offset < sizeOfHalfBlock; offset ++ ) {
45             @\tightcolorbox{custommagenta}{i1}@ = b0 | offset; // faster than, but equivalent to, b0+offset
46             @\textcolor{customblue}{if ( (i1 \& inclusionMask) != desiredValueMask )}@
47                 @\textcolor{customblue}{continue; // skip}@
48             @\tightcolorbox{customgreen}{i2}@ = @\tightcolorbox{custommagenta}{i1}@ | sizeOfHalfBlock; // equivalent to @\tightcolorbox{custommagenta}{i1}@+sizeOfHalfBlock
49             @\tightcolorbox{custommagenta}{b[i1]}@ = @\tightcolorbox{custommagenta}{U[0,0]}@*@\tightcolorbox{custommagenta}{a[i1]}@ + @\tightcolorbox{custommagenta}{U[0,1]}@*@\tightcolorbox{customgreen}{a[i2]}@;
50             @\tightcolorbox{customgreen}{b[i2]}@ = @\tightcolorbox{customgreen}{U[1,0]}@*@\tightcolorbox{custommagenta}{a[i1]}@ + @\tightcolorbox{customgreen}{U[1,1]}@*@\tightcolorbox{customgreen}{a[i2]}@;
51         }
52     }
53     return b;
54 }
\end{lstlisting}

In the code for \texttt{qubitWiseMultiply()}, lines 49 and 50 are the only ones involving complex arithmetic.

As with all pseudocode examples in this article,
the arrays are zero-based (i.e., an array \texttt{A} of length \texttt{L} has elements \texttt{A[0]} through \texttt{A[L-1]} \cite{dijkstra1982zeroBased}),
and the syntax for bitwise operators
(\texttt{\hspace{0pt}\char60\hspace{0pt}\char60\hspace{0pt}} for shift left, \texttt{|} for bitwise OR, \texttt{\char94} for bitwise XOR, \texttt{\&} for bitwise AND, \texttt{\char126} for bitwise NOT)
is the same as in C, C++, C\#, Java, JavaScript, and Python.
`For' loops are usually written as something like \texttt{for (i=0; i<L; i++) \char123...\char125 }, which is compatible with all these languages except Python,
where the equivalent loop could be written \texttt{for x in range(0,L,1):}

The parts of the pseudocode in blue process any control (or anti-control) qubits associated with the gate.
For example, to implement a $\mathrm{CX}_{j,k}$ gate with target qubit on wire $k$ and control qubit on wire $j$,
\texttt{qubitWiseMultiply()} would be called with arguments \texttt{i\_w=k}
and \texttt{listOfControlBits = [[j,true]]}.
To implement a Toffoli gate (also called CCX or CCNOT) with two control qubits $j_1$ and $j_2$,
we would pass in \texttt{listOfControlBits = [[j1,true], [j2,true]]}.

(Note that some software platforms do not support anti-control qubits,
requiring the use of a control qubit preceded and followed by an $X$ gate
to achieve the effect of an anti-control qubit.
However, such $X$ gates increase the depth of the circuit, and the computational cost of simulating the circuit.)


To apply this algorithm to the circuit in Figure~\ref{fig:exampleCircuit},
we must first rewrite the circuit so that each layer contains only one gate,
i.e., separating the $H_1$ and $X_2$ gates in the left-most layer to be in separate, consecutive layers.
Then we can simulate the circuit like this:

\begin{lstlisting}[style=pseudocode, basicstyle={\fontsize{7.9pt}{9.5pt}\selectfont\ttfamily\bfseries}]
n = 3; // number of qubits
@$\psi$@ = @$( | 0 \rangle^ {\otimes {\rm n}} )$@; // initialization
@$\psi$@ = qubitWiseMultiply(n,Hadamard,1,@$\psi$@);
@$\psi$@ = qubitWiseMultiply(n,PauliX,2,@$\psi$@);
@$\psi$@ = qubitWiseMultiply(n,PauliX,0,@$\psi$@,[[1,true]]);
@$\psi$@ = qubitWiseMultiply(n,PauliZ,0,@$\psi$@);
@$\psi$@ = qubitWiseMultiply(n,PauliX,2,@$\psi$@,[[1,true]]);
\end{lstlisting}

Each call to \texttt{qubitWiseMultiply()} takes $O(2^\NN)$ time.
More generally, consider a circuit on $\NN$ qubits with initial depth $\DD$,
and consider the worst case where each layer initially contains $\NN$ gates.
The circuit must be expanded to have $\NN \DD$ layers,
so that each layer now has one gate.
Applying qubit-wise multiplication
to each layer results in a total runtime of $O(2^\NN \NN \DD)$,
which is far less than the runtime of $O(4^\NN \DD)$ in the previous section.
In addition, qubit-wise multiplication easily allows for arbitrary combinations
of control and anti-control qubits on each gate.

\section{Efficient SWAP gate}

Here we explain how to efficiently simulate SWAP gates,
with optional, arbitrary combinations of control and anti-control qubits.

SWAP gates exchange the states of two qubits.
The effect of this on a state vector is to exchange amplitudes
whose indices in the vector correspond to bit strings with the associated bits exchanged.
For example, in a 5-qubit circuit, the state vector is 32$\times$1,
and the indices of its elements, in binary,
range from
\texttt{00000}
to
\texttt{11111},
with bit positions numbered 4 (left-most and most significant)
to 0 (right-most and least significant).
To SWAP qubits 0 and 3, for example,
each amplitude for index $b_4 b_3 b_2 b_1 b_0$
is swapped with the amplitude for index $b_4 b_0 b_2 b_1 b_3$.

Adding support for control qubits turns out to be a simple matter of
copying the blue portions of code from \texttt{qubitWiseMultiply()}.
The result is shown in the two subroutines below.

\begin{lstlisting}[style=pseudocode, basicstyle={\fontsize{5.8pt}{7.0pt}\selectfont\ttfamily\bfseries}]
01 // Returns the given number k with its ith and jth bits swapped.
02 // Bits are numbered from 0 for the least significant bit.
03 // Examples swapping first and last of 4 bits:
04 //   swapBits(14,0,3) returns 7, because 14==1110_2, 7==0111_2
05 //   swapBits(10,0,3) returns 3, because 10==1010_2, 3==0011_2
06 // Examples swapping the middle two of 4 bits:
07 //   swapBits(13,1,2) returns 11, because 13==1101_2, 11==1011_2
08 //   swapBits(10,1,2) returns 12, because 10==1010_2, 12==1100_2
09 swapBits( k, i, j ) {
10     if ( i==j ) return k;
11     bit_i = (k >> i) & 1;
12     bit_j = (k >> j) & 1;
13     if ( bit_i != bit_j ) {
14         mask = (1 << i) | (1 << j);
15         k ^= mask; // flip bits i and j
16     }
17     return k;
18 }
\end{lstlisting}

\begin{lstlisting}[style=pseudocode, basicstyle={\fontsize{5.3pt}{6.4pt}\selectfont\ttfamily\bfseries}]
20 // Returns the given state vector |a> after swapping wires i and j.
21 // In other words, implements a SWAP gate on qubits i and j.
22 // Takes O(2^n) time.
23 @\textcolor{customblue}{// Control bits and anti-control bits limit the effect of the SWAP }@
24 @\textcolor{customblue}{// to a subset of the amplitudes in |a>. }@
25 applySwap(
26     n,   // number of qubits in the circuit, 1 <= n
27     i_w, j_w, // indices of wires to swap, 0 <= i_w <= n-1, 0 <= j_w <= n-1
28
29     // This is the state vector to transform;
30     // a (2^n)@$\times$@1 column vector of complex amplitudes
31     a,
32 
33     @\textcolor{customblue}{// A list of pairs of the form [wire\_index, flag] where 0<=wire\_index<n}@
34     @\textcolor{customblue}{// and flag is true for a control bit and false for an anti-control bit}@
35     @\textcolor{customblue}{listOfControlBits = [ ] // empty by default}@
36 ) {
37     @\textcolor{customblue}{inclusionMask = 0;}@
38     @\textcolor{customblue}{desiredValueMask = 0;}@
39     @\textcolor{customblue}{for ( iterator : listOfControlBits ) \char123 }@
40         @\textcolor{customblue}{[ wireIndex, flag ] = iterator;}@
41         @\textcolor{customblue}{bit = 1 \hspace{0pt}\char60\hspace{0pt}\char60\hspace{0pt} wireIndex; // 2\char94 wireIndex }@
42         @\textcolor{customblue}{inclusionMask |= bit; // turn on the bit}@
43         @\textcolor{customblue}{if ( flag )}@
44             @\textcolor{customblue}{desiredValueMask |= bit; // turn on the bit}@
45     @\textcolor{customblue}{ \char125 }@
46 
47     sizeOfStateVector = 1 << n; // 2^n
48     b = a.copy(); // copies all amplitudes from a to b
49     if ( i_w == j_w ) return b; // there's no work to do
50     for ( k = 0; k < sizeOfStateVector; k ++ ) {
51         @\textcolor{customblue}{if ( (k \& inclusionMask) != desiredValueMask )}@
52             @\textcolor{customblue}{continue; // skip}@
53         k2 = swapBits( k, i_w, j_w );
54         if ( k2 > k ) { // this check ensures we don't swap each pair twice
55             // swap the (k)th and (k2)th amplitudes
56             b[k2] = a[k];
57             b[k] = a[k2];
58         }
59     }
60     return b;
61 }
\end{lstlisting}

The total runtime of \texttt{applySwap()} is $O(2^\NN)$.
Assuming no control (nor anti-control) bits are specified,
the call to \texttt{swapBits()} at line 53 will happen $2^\NN$ times, but line 54 will cause an actual swap to only happen $2^\NN/4$ times, because in 1/2 of the cases, the (\texttt{i\_w})th and (\texttt{j\_w})th bits are equal (both 0 or both 1), and in 1/4 of the cases,
they are different in such a way that \texttt{k2 < k}.
This produces correct output, but is wasteful in calls to \texttt{swapBits()}.
We can reduce the number of bitwise operations performed
by not using \texttt{swapBits()} at all,
and instead precomputing these before the loop at line 50:
\begin{lstlisting}[style=pseudocode, basicstyle={\fontsize{7.9pt}{9.5pt}\selectfont\ttfamily\bfseries}]
antimask_i = ~(1 << i_w);
mask_j = 1 << j_w;
\end{lstlisting}
and then modifying the code after line 52 to extract the (\texttt{i\_w})th bit,
and only if it is 1,
to then extract the (\texttt{j\_w})th bit,
and only if that is 0,
to then compute k2 and perform the swap:
\begin{lstlisting}[style=pseudocode, basicstyle={\fontsize{7.9pt}{9.5pt}\selectfont\ttfamily\bfseries}]
ithBitOfK = ( k >> i_w ) & 1;
if ( ithBitOfK==1 ) {
    jthBitOfK = ( k >> j_w ) & 1;
    if ( jthBitOfK==0 ) {
        // turn off bit i, and turn on bit j
        k2 = ( k & antimask_i ) | mask_j;
        // swap the (k)th and (k2)th amplitudes
        b[k2] = a[k];
        b[k] = a[k2];
    }
}
\end{lstlisting}
Now the code still takes $O(2^\NN)$ time but is faster,
possibly at the cost of being more difficult to understand.

\begin{figure}[!thb]
\centering
\includegraphics[width=0.87\columnwidth]{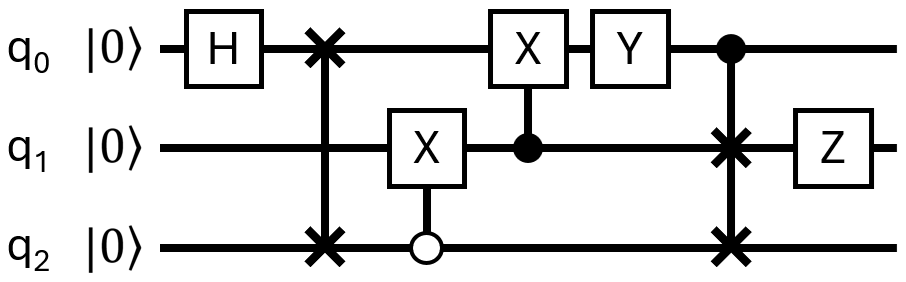}
\caption{An example circuit
   with one SWAP gate after the Hadamard,
   and one controlled SWAP (CSWAP) gate
   before the $Z$ gate.
   The output state vector is $(1/\sqrt{2})(i|010\rangle -i|011\rangle)$.}
\label{fig:swapExample}
\end{figure}


To illustrate how to use \texttt{applySwap()},
the circuit in Figure~\ref{fig:swapExample} would be simulated by doing
\begin{lstlisting}[style=pseudocode, basicstyle={\fontsize{7.8pt}{9.4pt}\selectfont\ttfamily\bfseries}]
n = 3; // number of qubits
@$\psi$@ = @$( | 0 \rangle^ {\otimes {\rm n}} )$@; // initialization
@$\psi$@ = qubitWiseMultiply(n,Hadamard,0,@$\psi$@);
@$\psi$@ = applySwap(n,0,2,@$\psi$@);
@$\psi$@ = qubitWiseMultiply(n,PauliX,1,@$\psi$@,[[2,false]]);
@$\psi$@ = qubitWiseMultiply(n,PauliX,0,@$\psi$@,[[1,true]]);
@$\psi$@ = qubitWiseMultiply(n,PauliY,0,@$\psi$@);
@$\psi$@ = applySwap(n,1,2,@$\psi$@,[[0,true]]);
@$\psi$@ = qubitWiseMultiply(n,PauliZ,1,@$\psi$@);
\end{lstlisting}

\section{Gates on two or more qubits}

The \texttt{qubitWiseMultiply()} and \texttt{applySwap()}
routines above make it easy to implement
certain gates acting on two or more qubits,
such as CX, Toffoli, and controlled SWAP (CSWAP or Fredkin) gates,
simply by passing in a list of control qubits.

However, there are other gates on two or more qubits,
like $i$SWAP or $\sqrt{{\rm SWAP}}$, that require more programming.
These gates have their own corresponding matrices, of size 4$\times$4 or greater.
To support them, it is straightforward to extend \texttt{qubitWiseMultiply()} to support
matrices of size greater than 2$\times$2.
This would then allow a gate $U$ on two or more qubits to be applied to {\em consecutive} qubits of a circuit.
To apply $U$ to qubits that are {\em not} consecutive,
we can write code that automatically applies one or more SWAP gates before $U$,
to temporarily ``rewire'' the relevant qubits to be consecutive, and then undo this rewiring after $U$ with one or more subsequent SWAPs.

\section{Analyzing Qubits}\label{sec:partialTrace}


Two popular web-based simulators for quantum circuits,
IBM Quantum Composer \cite{ibm2023composer} and Quirk \cite{gidney2020quirk},
not only simulate a circuit, but also display statistics about each of the qubits
at the output of the circuit:
the probability of measuring a 1,
and the phase and purity of each qubit, or the Bloch sphere coordinates of each qubit.
How are these statistics computed?

To find the probability that measuring a qubit results in a 1,
we can sum the probabilities for the amplitudes with indices where that qubit is 1.
For example, in a 3-qubit state vector, the state vector has eight amplitudes $a_i$, with index $i$ varying from \texttt{000} to \texttt{111} (in binary),
and the probability that qubit $q_2$ (for example) will yield a 1 when measured is
$|a_{100}|^2 + |a_{101}|^2 + |a_{110}|^2 + |a_{111}|^2$, i.e., an expression including all indices where bit 2 (the leftmost bit) is 1.
This probability can be computed for each of the 3 qubits.

Finding the phase, purity, or Bloch sphere coordinates of each qubit requires more work,
and involves finding a way to represent each qubit's state.
In general, it is not possible to find a $2 \times 1$ state vector that represents each qubit's state,
but it {\em is} possible to find a $2 \times 2$ {\em density matrix} for each qubit.
To give a high level summary of the process,
we start with the $2^\NN \times 1$ state vector $\ket{\psi}$ representing all $\NN$ qubits;
from that, we compute the $2^\NN \times 2^\NN$ density matrix $\rho = \ket{\psi}\bra{\psi}$;
from that, we use the {\em partial trace} to find the $2 \times 2$ {\em reduced density matrix} $\rho_i$ for each qubit;
and then from $\rho_i$
we can compute various statistics,
including phase, purity, Bloch sphere coordinates, and an alternative way to find the probability of measuring a 1.
The following subsections explain the concepts involved.


\subsection{Pure and Mixed States}

A {\em pure state} can be defined as a state where
everything there is to know about the state is known
(even if it involves a superposition).
Such a state can be described using a state vector.
A {\em mixed state}, however, is often defined as
a statistical ensemble of pure states, i.e.,
a combination of multiple pure states,
each with some classical probability.
Both kinds of states can be described with a density matrix.
In the context of a simulation of qubits with no noise,
where the initial state of the qubits is known perfectly,
we might guess that we won't need to deal with mixed states.
However, mixed states arise inevitably when we want to describe the state of a subset of qubits, or of a single qubit,
even if the complete set of $\NN$ qubits is in a pure state.

For concreteness,
consider the 8$\times$1 state vector $\ket{\psi_{210}}$
for all $\NN=3$ qubits of some circuit,
and let $\rho_{210} = \ketbra{\psi_{210}}{\psi_{210}}$ be the corresponding $8 \times 8$ density matrix.
(The 210 subscript indicates that these variables
pertain to all three qubits $q_2, q_1, q_0$).
Let $\rho_i$ be the $2\times 2$ density matrix for each
individual qubit $q_i$.
For our simulator, we assume that the state described by
$\rho_{210}$ is pure,
but each of the $\rho_i$ may be pure or mixed.
As mentioned already, each of the $\rho_i$ can be found
by performing an appropriate {\em partial trace} on $\rho_{210}$.

In the special case that each of the $\rho_i$ are pure,
then $\rho_{210}$ is a {\em product state},
and $\rho_{210} = \rho_2 \otimes \rho_1 \otimes \rho_0$,
and there exist $\ket{\psi_i}$
for $i = 0, 1, 2$
such that 
$\rho_i = \ketbra{\psi_i}{\psi_i}$
and $\ket{\psi_{210}} = \ket{\psi_2} \otimes \ket{\psi_1} \otimes \ket{\psi_0}$.
However, in general, $\rho_{210}$ is not a product state,
because there exist dependencies or correlations
between at least some of the qubits,
causing {\em entanglement}.

\subsection{Entangled States}

Entanglement is a property of a compound state
made up of two or more qubits.
The simplest definition of entanglement
is with respect to a bipartition
(i.e., a separation of a system into two subsystems $A$ and $B$,
or two subsets of qubits).
The partial trace is an essential tool in this context,
allowing us to describe one subsystem $A$ whilst the remaining complementary subsystem $B$ is ignored ({\em traced out}).
A bipartite pure state can either be a {\em product state} or an {\em entangled state},
while a bipartite mixed state can either be a product state, a {\em separable state} (generalizing a product state),
or an entangled state (sections 2.1-2.2 in \cite{guhne2009entanglement}).
The definition of entanglement can also be extended to multipartite systems.
The full details are beyond our scope,
but there are many resources for further reading
\cite{plenio2007introduction,guhne2009entanglement,horodecki2009quantum}.


\subsection{The Partial Trace Algorithm}

Partial trace is a generalization of the (standard) trace operation on a matrix which sums all diagonal elements,
yielding a single number.
With a {\em partial} trace, we start with a larger matrix,
and obtain a smaller matrix whose elements are each sums of elements taken from the larger matrix.
It is often defined formally as something like
$\Tr_B \left( \rho_{AB} \right) = \sum_{t=0}^{T-1} \left( \mathsf{I}_A \otimes \bra{t} \right)  \rho_{AB}  \left( \mathsf{I}_A \otimes \ket{t} \right)$,
where $A$ is the subsystem to keep and comprises $K$ qubits,
$B$ is the subsystem to trace out (or {\em trace over}) and comprises $T$ qubits,
$\rho_{AB}$ is a $(2^{K+T} \times 2^{K+T})$ density matrix,
$\mathsf{I}_A$ is the $(2^K \times 2^K)$ identity matrix,
and the $\ket{t}$ are basis states in the subspace of $B$
(for example, if $T = 3$, then $t$ could range from 000 to 111).
The result of $\Tr_B \left( \rho_{AB} \right)$
is a $(2^K \times 2^K)$ reduced density matrix.
Maziero \cite{maziero2017computing} discusses this definition and another commonly used (equivalent) definition,
as well as how to optimize the calculation,
but does not discuss how to trace out an arbitrary, non-adjacent subset of qubits.

Below, we provide pseudocode for a
\texttt{partialTrace()} routine that takes a list of qubits to trace out.
For example, if $\rho_{210}$ is 8$\times$8, we could invoke
\begin{lstlisting}[style=pseudocode, basicstyle={\fontsize{7.9pt}{9.5pt}\selectfont\ttfamily\bfseries}]
@$\rho_0$@ = partialTrace(3,@$\rho_{210}$@,[1,2]);
@$\rho_1$@ = partialTrace(3,@$\rho_{210}$@,[0,2]);
@$\rho_2$@ = partialTrace(3,@$\rho_{210}$@,[0,1]);
\end{lstlisting}
to obtain the 2$\times$2 reduced density matrix for each qubit.
Each of these partial traces corresponds to a different bipartition of the original state $\rho_{210}$.
On the left-hand-side of the assignment statements above, the subscripts of the $\rho$ matrices indicate the subset of qubits retained after tracing out the other qubits.


Taking things to an extreme,
tracing out all qubits with \texttt{partialTrace(3,$\rho_{210}$,[0,1,2])} results in a standard trace,
i.e., a 1$\times$1 matrix containing the sum of only the diagonal elements,
which is always equal to 1 if we are tracing a density matrix.



\begin{lstlisting}[style=pseudocode, basicstyle={\fontsize{4.9pt}{5.9pt}\selectfont\ttfamily\bfseries}]
01 // Returns the given number i with its bits rearranged,
02 // so that the kth bit of i is returned in position a[k]. Examples:
03 //   rearrangeBits(i,[1,0]) returns the two least-significant bits of i,
04 //     swapped, and none of the other bits.
05 //   rearrangeBits(i,[0,1,2])) returns only the three least-significant bits of i,
06 //     with their positions unchanged.
07 //   rearrangeBits(i,[3,0,1,2])) returns only the four least-significant bits of i,
08 //     shifted left (to one position more significant) and wrapped around.
09 rearrangeBits( i, a /* an array of new positions */ ) {
10     returnValue = 0;
11     for ( position = 0; position < a.length; position ++ ) {
12         if ( a[position] >= 0 )
13             returnValue |= ( (i >> position) & 1 ) << a[position];
14     }
15     return returnValue;
16 }
\end{lstlisting}


\begin{lstlisting}[style=pseudocode, basicstyle={\fontsize{4.3pt}{5.2pt}\selectfont\ttfamily\bfseries}]
18 // Consider a 16@$\times$@16 density matrix M defined for 4 qubits, numbered 0 to 3.
19 // The caller can invoke partialTrace( 4, M, [0,2] );
20 // to trace out qubits 0 and 2, keeping 1 and 3, and returning a 4@$\times$@4 matrix.
21 // If T is the number of qubits to trace out, and K=n-T is the number of qubits to keep,
22 // the routine returns a matrix of size (2^K)@$\times$@(2^K) and takes O( (2^T) (2^(2K)) K ) time.
23 partialTrace(
24     n, // number of qubits in the circuit
25     inputMatrix, // a (2^n)@$\times$@(2^n) matrix of complex numbers
26     
27     // An array of values in the range 0 to n-1, representing the qubits to trace out.
28     // Assumed to be in ascending order and without duplicates.
29     qubitsToTraceOut
30 ) {
31     // Compute an array of complementary indices called qubitsToKeep,
32     // containing all the indices in [0,n-1] that are not in qubitsToTraceOut.
33     isTracedOut = [ ]; // this is a temporary array, initially empty
34     for ( i = 0; i < n; i++ ) // append n false values
35         isTracedOut.push( false );
36     for ( i = 0; i < qubitsToTraceOut.length; i++ )
37         isTracedOut[ qubitsToTraceOut[i] ] = true;
38     // Now, isTracedOut[i]==true means qubit i will be traced out
39     qubitsToKeep = [ ]; // the marginal qubits
40     for ( i = 0; i < n; i++ )
41         if ( ! isTracedOut[i] )
42             qubitsToKeep.push( i );
43 
44     numQubitsToTraceOut = qubitsToTraceOut.length;
45     numQubitsToKeep = qubitsToKeep.length;
46     assert( numQubitsToTraceOut + numQubitsToKeep == n ); // sanity check
47     // This is 2^numQubitsToTraceOut == the dimension of the space being traced out
48     tracedDimension = 1 << numQubitsToTraceOut;
49     // This is 2^numQubitsToKeep == the dimension of the resulting matrix
50     resultDimension = 1 << numQubitsToKeep;
51 
52     outputMatrix = new Matrix( resultDimension, resultDimension ); // initialized with zeros
53     for (
54         shared_bits = 0; // bits common to input_row and input_col
55         shared_bits < tracedDimension;
56         shared_bits ++
57     ) {
58         shared_bits_rearranged = rearrangeBits( shared_bits, qubitsToTraceOut );
59         for ( output_row = 0; output_row < resultDimension; output_row ++ ) {
60             input_row = shared_bits_rearranged | rearrangeBits( output_row, qubitsToKeep );
61             for ( output_col = 0; output_col < resultDimension; output_col ++ ) {
62                 input_col = shared_bits_rearranged | rearrangeBits( output_col, qubitsToKeep );
63                 outputMatrix[output_row,output_col] += inputMatrix[input_row,input_col];
64             }
65         }
66     }
67     return outputMatrix;
68 }
\end{lstlisting}

In the code for \texttt{partialTrace()}, line 63 is the only one involving complex arithmetic.

\subsection{Gaining Intuition for Partial Trace}

\begin{figure}[!thb]
 \centering
  \ifthenelse{\boolean{UseHighResolutionImages}}{
    \includegraphics[width=0.6\columnwidth]{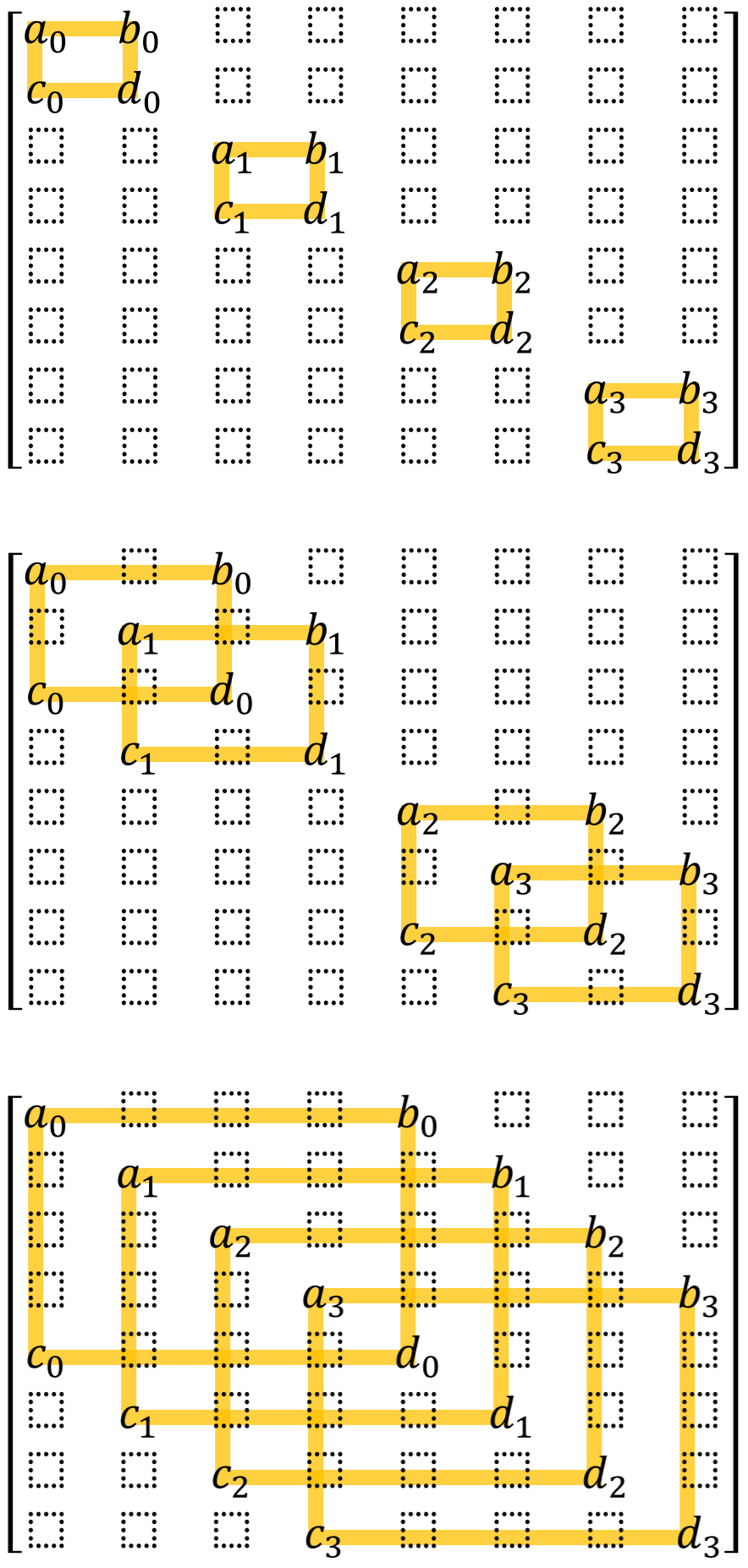}
  }{
    \includegraphics[width=0.6\columnwidth]{fig/partialTrace-v2.png}
  } 
  \caption{
     A partial trace applied to an 8$\times$8 matrix, to trace out two qubits,
     picks out certain elements (labeled $a_i, \ldots, d_i$ above) to sum,
     producing a 2$\times$2 result
     $\scalebox{0.8}{$\left[ \begin{array}{@{}cc@{}} A & B  \\ C & D   \end{array} \right]$}$
     where $A = \Sigma_i a_i$, ..., $D = \Sigma_i d_i$.
     If the 8$\times$8 matrix is a density matrix $\rho_{210}$,
     with rows and columns numbered 0 through 7,
     then above we see the elements picked out by
     $\mathrm{Tr_{21}}(\rho_{210})$ (top),
     $\mathrm{Tr_{20}}(\rho_{210})$ (middle),
     and $\mathrm{Tr_{10}}(\rho_{210})$ (bottom),
     where $\mathrm{Tr_{ij}}$ denotes tracing out bits $i$ and $j$.
     In other words, the labelings illustrate calls to \texttt{partialTrace()}
     with \texttt{qubitsToTraceOut} set to
     \texttt{[1,2]} (top),
     \texttt{[0,2]} (middle),
     \texttt{[0,1]} (bottom).
     Each orange rectangle corresponds to an iteration of the loop at line 53 in the \texttt{partialTrace()} pseudocode,
     and the subscripts on $a_i, \ldots, d_i$ are values of \texttt{shared\_bits}.
     Compare with the patterns in Figure~\ref{fig:qubitWise}.
  }
 \label{fig:partialTrace}
\end{figure}

Figure~\ref{fig:partialTrace} shows elements of an 8$\times$8 matrix
that are summed together by partial traces.
Notice that the orange rectangles in that figure are always
located on a diagonal;
this is related to the fact that the \texttt{shared\_bits} are used
in both the \texttt{input\_row} and \texttt{input\_col}
indices.

As another example, consider a 32$\times$32 density matrix,
on which we perform a partial trace, to trace out qubits
1, 2, and 4, keeping qubits 0 and 3.
Because three qubits are being traced out,
\texttt{shared\_bits} will vary from \texttt{000} to \texttt{111},
in binary.
Let the string $s_2 s_1 s_0$
denote the value of \texttt{shared\_bits},
where $s_0$ is the least significant bit.
\texttt{shared\_bits\_rearranged} will be a 5-bit value of the form
$s_2 0 s_1 s_0 0$, computed at line 58,
with the bits of interest at positions 1, 2, and 4
(the bits being traced out).
Because two qubits are being kept,
\texttt{output\_row} and \texttt{output\_col} each vary from
\texttt{00} to \texttt{11}, in binary.
Let $r_1 r_0$ and $c_1 c_0$ denote their values.
When rearranged (on lines 60 and 62, respectively),
they have the form $0 r_1 0 0 r_0$ and $0 c_1 0 0 c_0$,
with the bits of interest at positions 0 and 3
(the bits being kept).
Finally, these are ``or''d together with \texttt{shared\_bits\_rearranged}
to produce (\texttt{input\_row}, \texttt{input\_col}) =
($s_2 r_1 s_1 s_0 r_0$,
$s_2 c_1 s_1 s_0 c_0$),
which are used to index an element in the input matrix.
The $s$ bits always have the same value in the row and column,
making the orange rectangles in Figure~\ref{fig:partialTrace} fall along a diagonal,
but the $r$ and $c$ bits have independent values,
causing them to access the different positions
within each orange rectangle.

It is no accident that the orange rectangles in Figure~\ref{fig:partialTrace}
match those in Figure~\ref{fig:qubitWise}.
If a partial trace is performed on a matrix representing a product state
(i.e., a state equal to a tensor product of states),
then the partial trace is a way to ``undo'' that tensor product.
Another way to say this is as follows.
If $\rho_A$ and $\rho_B$ are two square matrices,
then $\mathrm{Tr_B}(\rho_A \otimes \rho_B) = \rho_A\,\Tr(\rho_B)$,
where $\mathrm{Tr_B}$ denotes tracing out the bits associated with $\rho_B$,
or tracing out the subsystem $B$.
If, furthermore, $\rho_B$ is a density matrix, then $\Tr(\rho_B) = 1$, and we have
$\mathrm{Tr_B}(\rho_A \otimes \rho_B) = \rho_A$.

\subsection{Statistics Describing Individual Qubits}

Stepping back from the details of the \texttt{partialTrace()}
algorithm,
once we have the 2$\times$2 density matrix $\rho_i$ for each $i$th qubit in a circuit,
we can compute statistics about each qubit.
It turns out that every 2$\times$2 density matrix is of the form
\begin{equation}
    \rho_i = (I + x X + y Y + z Z)/2 \nonumber
\end{equation}
where $I$, $X$, $Y$, $Z$ are the identity and Pauli matrices defined earlier,
and $x$, $y$, $z$ are the real-valued coordinates
in the Bloch sphere (Figure~\ref{fig:bloch})\footnote{In fact, there is a vector space of 2$\times$2 Hermitian matrices spanned by
$\{I, X, Y, Z\}$, and this is sometimes called the {\em space of density operators}. 
The Bloch sphere is embedded in a 3D projection of that space.}
associated with the qubit's state.
Keeping in mind that $\rho_i$'s diagonal elements are real and sum to 1
and its off-diagonal elements are conjugates of each other,
we can set $\rho_i = \left[ \begin{array}{@{}cc@{}} a & b+ic  \\ b-ic & 1-a \end{array} \right]$,
and then rewrite the above matrix equation as scalar equations,
and solve to find $x=2b$, $y=-2c$, $z=2a-1$.
We arrive at the same outcome using the facts that
$x=\Tr(\rho_i X)$, $y=\Tr(\rho_i Y)$, $z=\Tr(\rho_i Z)$.
Given $(x,y,z)$, we can then find
the radius $r = \sqrt{x^2+y^2+z^2}$
and angles $\theta$, $\phi$ shown in Figure~\ref{fig:bloch}.
The qubit's state is either pure or mixed,
resulting in $r = 1$ (on the Bloch sphere's surface)
or $r < 1$ (in the interior), respectively\footnote{This
helps to explain why the density matrix for a mixed state can be expressed as convex combination of pure states.}.
The qubit's phase is $\theta$,
and its purity $p = \Tr(\rho_i^2)$
is given by the trace of the squared density matrix,
where $p$ is either equal to 1 for a pure state,
or $0.5 \leq p < 1$ for a mixed state.
(More generally, for a density matrix of size $2^\NN \times 2^\NN$,
purity ranges from $1/2^\NN$ to 1.)
Earlier, we showed one way to find the
probability of measuring 1 on the qubit,
but this probability is also simply the lower right element of $\rho_i$,
which, it is interesting to notice, is easy to interpret as a
vertical position in the Bloch sphere (that probability being $1-a = (1-z)/2$).
Another statistic we might compute is the
linear entropy, defined as $1 - p$,
which is a metric of {\em mixedness}, the opposite of purity.


\subsection{Statistics Describing Pairs of Qubits}

There are also statistics we might compute
for a pair of qubits of interest.
As an example,
in a set of $\NN=5$ qubits,
to better understand the relationship between, say, qubits $q_1$ and $q_3$,
we could compute the 32$\times$32 density matrix
$\rho = | \psi \rangle \langle \psi |$ for all 5 qubits,
and then compute the 4$\times$4 reduced density matrix
for qubits $q_1$ and $q_3$ with
\begin{lstlisting}[style=pseudocode, basicstyle={\fontsize{7.9pt}{9.5pt}\selectfont\ttfamily\bfseries}]
@$\rho_{31}$@ = partialTrace(@$\NN$@,@$\rho$@,[0,2,4]);
\end{lstlisting}
From this, we can compute
several statistics about the pair of qubits,
such as:
the purity $p = \Tr((\rho_{31})^2)$ of the 4$\times$4 matrix,
where $1/4 \leq p \leq 1$;
the linear entropy $1 - p$;
the concurrence of the two qubits (see equations 1 and 2 in \cite{coffman2000distributed}) which is a metric of
entanglement
(one of many possible metrics of entanglement \cite{plenio2007introduction,guhne2009entanglement,horodecki2009quantum});
and the von Neumann entropy (see equation 11.40 in \cite{nielsen2010})
which is another metric of the mixedness of the 4$\times$4 matrix.

\subsection{Examples of Pure and Mixed States}

\begin{figure}[!thb]
\centering
\includegraphics[width=0.39\columnwidth]{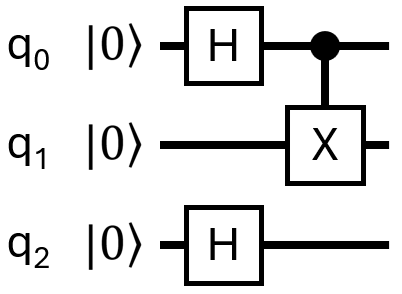}
\caption{The output state vector of this circuit
   is $\ket{\psi_f} = (1/2)(\ket{000}+\ket{011}+\ket{100}+\ket{111})$.
   We use this to perform some examples of partial traces.}
\label{fig:exampleCircuitForPartialTrace}
\end{figure}

Figure~\ref{fig:exampleCircuitForPartialTrace} shows another example circuit.
The reader can use the output state $\ket{\psi_f}$ of that circuit
to compute the 8$\times$8 density matrix $\rho_{210} = \ket{\psi_f}\bra{\psi_f}$,
and from that, compute various partial traces.
For example, 
the reader can compute these two states
\begin{align*}
\rho_{2} = \mathrm{Tr_{10}}(\rho_{210}) = \scalebox{0.6}{$\left[ \begin{array}{@{}cc@{}} 0.5 & 0.5 \\ 0.5 & 0.5 \end{array} \right]$}, && \rho_{10} = \mathrm{Tr_{2}}(\rho_{210}) = \scalebox{0.6}{$\left[ \begin{array}{@{}cccc@{}} 0.5 & 0 & 0 & 0.5  \\ 0 & 0 & 0 & 0 \\ 0 & 0 & 0 & 0  \\ 0.5 & 0 & 0 & 0.5 \end{array} \right]$}
\end{align*}
which turn out to be both pure,
as can be confirmed by checking that the trace of the square of each matrix is equal to 1
($\Tr((\rho_{2})^2) = 1$ and $\Tr((\rho_{10})^2) = 1$).
It makes sense that the qubit subsets $\{2\}$ and $\{1,0\}$ are each pure
because there is no entangling gate acting across them
in Figure~\ref{fig:exampleCircuitForPartialTrace}.

On the other hand,
these two states
\begin{align*}
\rho_{0} = \mathrm{Tr_{21}}(\rho_{210}) = \scalebox{0.6}{$\left[ \begin{array}{@{}cc@{}} 0.5 & 0 \\ 0 & 0.5 \end{array} \right]$}, && \rho_{21} = \mathrm{Tr_{0}}(\rho_{210}) = \scalebox{0.45}{$\left[ \begin{array}{@{}cccc@{}} 0.25 & 0 & 0.25 & 0  \\ 0 & 0.25 & 0 & 0.25 \\ 0.25 & 0 & 0.25 & 0  \\ 0 & 0.25 & 0 & 0.25 \end{array} \right]$}
\end{align*}
are both mixed,
as can be confirmed by checking that the trace of the square of each matrix is less than 1.
This reflects the fact that there {\em is} an entangling gate (CX) acting across subsets $\{0\}$ and $\{2,1\}$.

\subsection{Complexity of Partial Trace}

To analyze the complexity of
\texttt{partialTrace()},
let $\NN$ be the total number of qubits,
of which $T$ are to be traced out,
and $K = \NN - T$ are to be kept.
The \texttt{for} loops at lines 53, 59, 61
iterate $2^T$, $2^K$, and $2^K$ times, respectively.
In the innermost loop, each call to \texttt{rearrangeBits()}
takes $O(K)$ time.
Multiplying these, the total time for \texttt{partialTrace()} is
$O( 2^T 4^K K )$.
The next section shows how to improve this.

\subsection{Improvements to the Partial Trace Algorithm}

Several improvements to the algorithm are possible.
First, rather than calling \texttt{rearrangeBits()}
inside the two innermost loops (lines 60 and 62),
we could construct a lookup table by doing this before line 53:
\begin{lstlisting}[style=pseudocode, basicstyle={\fontsize{6.5pt}{7.8pt}\selectfont\ttfamily\bfseries}]
lookupTable = [];
for ( tmp = 0; tmp < resultDimension; tmp ++ ) {
    lookupTable[tmp] = rearrangeBits( tmp, qubitsToKeep );
}
\end{lstlisting}
and then modifying lines 60 and 62 to use the lookup table rather than
calling \texttt{rearrangeBits()}.
This would reduce the total runtime to $O( 2^T 4^K )$,
equivalent to the runtime in \cite{maziero2017computing}.

Second, if we assume that the input matrix passed to \texttt{partialTrace()}
is always a density matrix, then we know that the input and output matrices are both Hermitian.
Hence, the innermost loop (line 61) can be modified to only traverse
one triangular half (and the diagonal) of the output matrix,
e.g.,
\begin{lstlisting}[style=pseudocode, basicstyle={\fontsize{6.5pt}{7.8pt}\selectfont\ttfamily\bfseries}]
for ( output_col = 0; output_col <= output_row; output_col ++) {
    ...
}
\end{lstlisting}
After the outermost loop (line 53) terminates,
a separate pass over the output matrix can copy entries from the 1st triangular half into the 2nd triangular half, conjugating entries as they are copied.
This would reduce the total runtime by almost half.
Additional changes could define a special object for storing Hermitian
matrices, where the object only explicitly stores
one triangular half (plus the diagonal) of the matrix.
Using this special object to store the input and output matrices
would reduce by almost half the memory used.

Our last suggestion related to performance yields, by far,
the greatest improvement.
If the client must first
compute the full density matrix $\rho$ from a state vector
by computing $\rho = | \psi \rangle \langle \psi |$,
and then call \texttt{partialTrace()},
then the cost of doing both is dominated by constructing $\rho$,
and requires $O(4^\NN)$ time and $O(4^\NN)$ memory.
Figure~\ref{fig:partialTrace} hints that \texttt{partialTrace()} will only sparsely read elements in $\rho$,
hence it may not be necessary to compute all of $\rho$.
We can save a great amount of time and memory by modifying
\texttt{partialTrace()} to accept $| \psi \rangle$ as input instead of $\rho$,
and modifying line 63 to compute the required element of $\rho$ on demand with
\begin{lstlisting}[style=pseudocode, basicstyle={\fontsize{6.5pt}{7.8pt}\selectfont\ttfamily\bfseries}]
outputMatrix[output_row,output_col]
    += psi[input_row] * psi[input_col].conjugate();
\end{lstlisting}
In this way, the client need never compute the full $\rho$ explicitly.
To give some idea of the difference in performance this can make,
we generated random state vectors (using Muqcs \cite{mcguffin2025muqcs} running inside Chrome),
and measured the time to compute the full density matrix,
and the time to perform partial traces,
using either the full density matrix or the state vector as input. Here are some results:
\noindent\includegraphics[width=0.99\columnwidth]{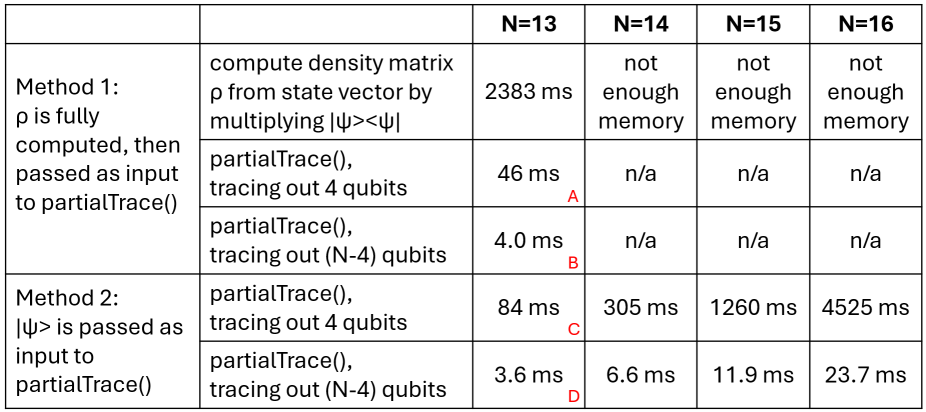}
It is clear that computing the full density matrix makes method 1 far more expensive.
Comparing individual numbers,
the time marked C in the results is greater than A, as expected, since C involves computing
entries of $\rho$ on demand rather than looking them up.
We were surprised that D is smaller than B, and suspect this is because B involves
accessing such a large $\rho$ ($(2^{13})^2$ $\times$ 16 bytes per complex number = 1 gigabyte)
leading to many cache misses.

Another possible improvement, related to convenience rather than performance,
is motivated by noticing that often,
the caller to \texttt{partialTrace()}
wants to trace out all except one or two qubits.
Thus, rather than passing in a long list of \texttt{qubitsToTraceOut},
it would be more elegant if the caller could pass in a
generic list of qubits,
along with a boolean flag to specify how to interpret that list:
as an enumeration of qubits to trace out, or qubits to keep.
To trace out all but one qubit, the caller could simply pass in a list containing that one qubit, with the appropriate flag value.




\section{Measurement Gates}

Measurement gates are different from the other gates discussed so far.
They are not unitary, and their effect cannot be fully modeled by simply
updating a state vector.
Given a state vector for $\NN$ qubits prior to any measurement,
the outcome of a measurement gate can be modeled by two new state vectors,
each half the size of the previous state vector,
corresponding to a measurement of 0 or 1, respectively.
As a concrete example, consider 3 qubits prior to any measurement,
and their 8$\times$1 state vector $| \psi \rangle = a_{000}|000\rangle + \ldots + a_{111}|111\rangle$.
A measurement gate applied to qubit $q_2$
results in a 0 with probability
$\Pr[q_2=0] = |a_{000}|^2 + |a_{001}|^2 + |a_{010}|^2 + |a_{011}|^2$,
and a 1 with probability
$\Pr[q_2=1] = |a_{100}|^2 + |a_{101}|^2 + |a_{110}|^2 + |a_{111}|^2$.
For each of these outcomes, the remaining qubits are modeled
with a new, smaller state vector whose amplitudes are normalized by the probabilities,
specifically
\begin{equation}
    |\psi_{q_2=0}\rangle = \frac{a_{000}|00\rangle + a_{001}|01\rangle + a_{010}|10\rangle + a_{011}|11\rangle}{\sqrt{\Pr[q_2=0]}} \nonumber
\end{equation}

\begin{equation}
  |\psi_{q_2=1}\rangle = \frac{a_{100}|00\rangle + a_{101}|01\rangle + a_{110}|10\rangle + a_{111}|11\rangle}{\sqrt{\Pr[q_2=1]}} \nonumber
\end{equation}

In our discussion, we assume that measurement is performed in the computational basis
(i.e., along the $z$ axis in the Bloch sphere),
however it is often useful to measure in other bases,
which can be achieved with an appropriate rotation in the Bloch sphere.
For example, measuring along the $x$ axis is equivalent
to applying a Hadamard gate and then measuring along $z$.

There are several ways to implement measurement gates in a simulator.
First, the software could randomly choose one measurement outcome
(where the random choice is weighted by the probabilities)
and continue simulating with the now smaller state vector.
The simulation of the entire circuit can then be placed inside a repeating
loop for, say, 100 or 1000 iterations,
while the software collects statistics on the outcomes of these iterations.
This would resemble the actual functioning of a quantum computer.

Second, the software could allow the user to interactively choose which
measurement outcome to simulate, allowing the user to manually explore
each branch into the future.

Third, the software could store both outcomes, with their probabilities and smaller state vectors,
simulating each branch into the future.
If there are multiple measurement gates, this results in a tree of possibilities.
If there are $\NN$ qubits and $\MM$ measurement gates encountered so far,
this results in $2^\MM$ state vectors each of size $2^{\NN-\MM} \times 1$,
requiring $O(2^\NN)$ memory.
This third approach would allow for precise calculation of expected final probabilities
in just one simulation of the circuit.

Fourth, the software could store a density matrix
$\rho = \Pr[q_2=0] |\psi_{q_2=0}\rangle \langle \psi_{q_2=0}|
      + \Pr[q_2=1] |\psi_{q_2=1}\rangle \langle \psi_{q_2=1}|$.
The density matrix $\rho$ represents a mixed state, and is equal to a weighted sum of pure states.
(In general, the density matrix for any mixed state can be decomposed into a weighted
sum of density matrices of pure states, where the weights are probabilities.)
The simulator can then update the density matrix under the effect of any subsequent
unitary layer $L_j$ of the circuit by using the update rule $\rho_{j+1} = L_j \rho_j L_j^\HT$
(this matrix product can be computed more efficiently
using a variant of qubit-wise multiplication
applied column-by-column and row-by-row,
but is still much more expensive than updating a state vector).
If there are $\NN$ qubits and $\MM$ measurement gates encountered so far,
this results in one density matrix of size $2^{\NN-\MM} \times 2^{\NN-\MM}$,
requiring $O(4^{\NN-\MM})$ memory.
The worst case occurs when we encounter the first measurement gate,
requiring $O(4^{\NN-1})$ memory.
Thus, this 4th approach is expensive in time and memory,
and also has the disadvantage that there is no guarantee that the density matrix
can be uniquely decomposed into pure states.

A fifth approach is to use the deferred measurement principle (section 4.4 in \cite{nielsen2010}) to move all measurement gates to the end of the circuit.
All previous layers in the circuit are then unitary,
and can be simulated using a single state vector.
This is simple and efficient, but depending on the circuit, it may not always be convenient,
and may not be desirable
if the users wishes to analyze and understand different branches of possible measurement outcomes
with the measurement gates appearing earlier in the circuit.

The first and fifth approaches are probably the simplest to implement,
although the fifth is not always appropriate.
The second and third make it possible for a user to interactively
examine different branches of execution.
The fourth approach scales more poorly than the other approaches.

\section{Further Enhancements}

The subroutines in this tutorial could be made more robust
by adding error checking,
e.g., doing bounds checks on arguments passed in,
ensuring that \texttt{listOfControlBits} does not contain contradictory entries where a qubit is both control and anti-control,
ensuring that the \texttt{qubitsToTraceOut} passed in to \texttt{partialTrace()}
are in ascending order and without duplicates,
etc.
To save memory, the subroutines could also be modified to perform all updates to the state vector
in place, rather than allocating and returning a new state vector.

A large speedup could be achieved using GPU programming \cite{heng2020,zhang2021hyquas,zhao2022}.
A mid-range Nvidia GPU chip contains thousands of CUDA cores
and has high-speed access to enough memory to store a state vector for 20-30 qubits.

Other kinds of classical simulators can also be implemented \cite{xu2023}
depending on the kind of circuit.
For example, tensor networks \cite{markov2008} \cite[section 6.3]{viamontes2009qcsim} \cite{arad2010,vaneersel2010tensor,biamonte2017,bridgeman2017,taylor2024tensor,berezutskii2025}
can be used to accelerate computations.
As another example,
if the circuit is limited to Clifford operations
(operations that can be generated by composing $H$, $S = Z^{0.5}$, and CX,
which includes $X$, $Y$, $Z$, $X^{0.5}$, $Y^{0.5}$, SWAP, $i$SWAP,
but not $T = Z^{0.25}$),
then the Gottesman-Knill algorithm \cite{anders2006stabilizer} can simulate the circuit in polynomial time.

\section{Conclusions}

Developing a simulator from scratch yields insights beyond what can be gained from using existing software.
This tutorial has presented efficient algorithms for core operations enabling simulation of circuits on a laptop,
to lower the barrier to entry for students and practitioners seeking a deeper understanding of quantum computing.



%



\ifthenelse{\boolean{IncludeAppendix}}{

\appendix[Calibration of HoloLens]



We implemented calibration with a homography
matrix that transforms ...

}{} 

\ifCLASSOPTIONcompsoc
  \section*{Acknowledgments}
\else
  \section*{Acknowledgment}
\fi

This research was supported by NSERC (MJM), and by the NSF under Grant No. OSI-2328774 (KI).

\ifCLASSOPTIONcaptionsoff
  \newpage
\fi




\bibliography{main}
\bibliographystyle{IEEEtran}

%


\begin{IEEEbiography}[{\includegraphics[width=1in]{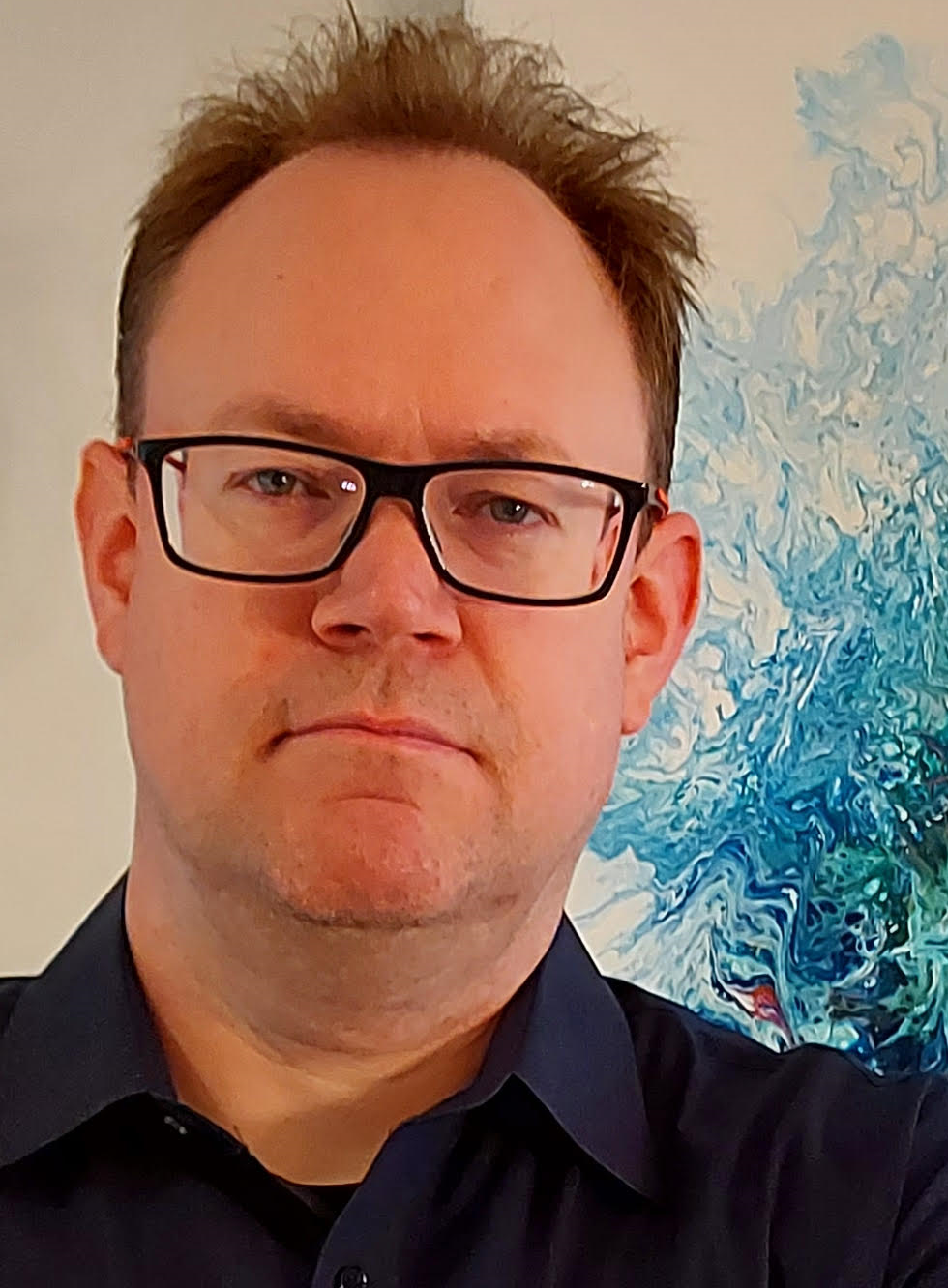}}]{Michael J. McGuffin} is a full professor at ETS, a French-language engineering school in Montreal, Canada, where his students do research in HCI and visualization.
His recent interests include quantum computing, visual programming, and artificial life.
In 2009, his paper at the IEEE Information Visualization Conference (InfoVis 2009) received an Honorable Mention.
\end{IEEEbiography}

\begin{IEEEbiography}[{\includegraphics[width=1in]{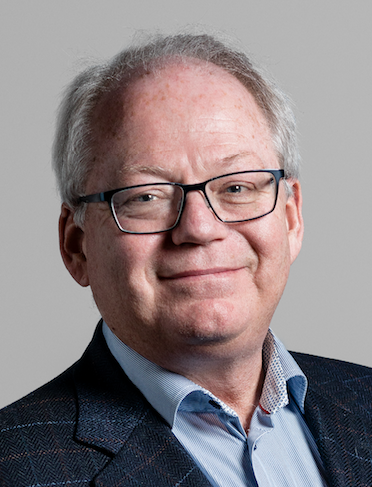}}]{Jean-Marc Robert} is a full professor at ETS
where his students do research in IT security and quantum cryptography. His recent interests include quantum computing and quantum information theory. In 1985, he completed his Master's degree, formalizing the concept of privacy amplification, which led to the second paper ever published on quantum cryptography.
\end{IEEEbiography}

\begin{IEEEbiography}[{\includegraphics[width=1in]{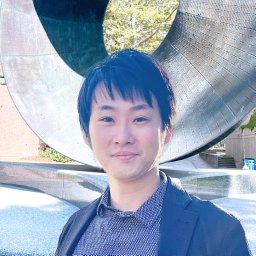}}]{Kazuki Ikeda} is an Assistant Professor at the University of Massachusetts Boston.
His website is \url{https://kazukiikeda.studio.site/}
\end{IEEEbiography}







\end{document}